\documentclass[%
 reprint,%
 amssymb, amsmath,%
 aip,cha,%
]{revtex4-1}
\usepackage{graphicx}
\usepackage{amsmath}

\begin{document}

\title{Diffusion-driven growth of nanowires by low-temperature molecular beam epitaxy.}

\author{P. Rueda-Fonseca}
\author{M. Orr\`{u}}
\affiliation{Univ. Grenoble Alpes, F-38000 Grenoble, France}
\affiliation{CNRS, Institut NEEL, F-38000 Grenoble, France}
\affiliation{CEA, INAC, F-38000 Grenoble, France}

\author{E. Bellet-Amalric}
\author{E. Robin}
\affiliation{Univ. Grenoble Alpes, F-38000 Grenoble, France}
\affiliation{CEA, INAC, F-38000 Grenoble, France}

\author{M. Den Hertog} \affiliation{Univ. Grenoble Alpes, F-38000 Grenoble,
France} \affiliation{CNRS, Institut NEEL, F-38000 Grenoble, France}

\author{Y. Genuist} \affiliation{Univ. Grenoble Alpes, F-38000 Grenoble, France} \affiliation{CNRS, Institut NEEL, F-38000 Grenoble, France}

\author{R. Andr\'e} \affiliation{Univ. Grenoble Alpes, F-38000 Grenoble,
France} \affiliation{CNRS, Institut NEEL, F-38000 Grenoble, France}

\author{S. Tatarenko} \affiliation{Univ. Grenoble Alpes, F-38000 Grenoble,
France} \affiliation{CNRS, Institut NEEL, F-38000 Grenoble, France}

\author{J. Cibert} \email{joel.cibert@neel.cnrs.fr} \affiliation{Univ. Grenoble Alpes, F-38000 Grenoble,
France} \affiliation{CNRS, Institut NEEL, F-38000 Grenoble, France}
\date{\today{}}

\date{\today}

\begin{abstract}
With ZnTe as an example, we use two different methods to unravel the
characteristics of the growth of nanowires by gold-catalyzed
molecular beam epitaxy at low temperature. In the first approach,
CdTe insertions have been used as markers, and the nanowires have
been characterized by scanning transmission electron microscopy,
including geometrical phase analysis, and energy dispersive electron
spectrometry; the second approach uses scanning electron microscopy
and the statistics of the relationship between the length of the
tapered nanowires and their base diameter. Axial and radial growth
are quantified using a diffusion-limited model adapted to the growth
conditions; analytical expressions describe well the relationship
between the NW length and the total molecular flux (taking into
account the orientation of the effusion cells), and the
catalyst-nanowire contact area. A long incubation time is observed.
This analysis allows us to assess the evolution of the diffusion
lengths on the substrate and along the nanowire sidewalls, as a
function of temperature and deviation from stoichiometric flux.

\end{abstract}

\maketitle

\section{introduction}

The  core-shell nanowire (NW) configuration is extremely flexible
and virtually any material combination, with any mismatch (up to a
critical thickness which can be large)\cite{Nazarenko} or no
mismatch, can be realized. A first, obvious application in
semiconductor materials is the fabrication of a shell with a larger
bandgap in order to reduce the effect of surface defects on the
properties of carriers confined in the core.\cite{demichel,joyce} A
tapered (cone-shaped) configuration may even be tailored to improve
light extraction.\cite{claudon} A natural extension is a dot
embedded in a NW, where the shape of the insertion controls the
confinement of the carriers - electrons and holes. Many
opportunities are then opened for a precise engineering of the
envelope functions, not only through the natural band offsets, but
also through a proper design of the built-in strain and the
associated piezoelectric field and deformation
potential.\cite{Niquet, Boxberg, Ferrand} In particular, it allows
one to engineer the hole states in a semiconductor dot and tailor
their orbital and spin states: a flat insertion induces a strain
configuration (hence a light-hole heavy-hole splitting) quite
similar to that of a dot made of the same materials but grown by the
Stranski-Krastanov mechanism, while the configuration is reversed in
a core-shell NW\cite{Ferrand} or an elongated dot. This opens the
opportunity to fully design the photonic properties of the
nanostructure, and its magnetic properties if magnetic impurities
are added.

To grow a well-controlled core-shell structure, two different
strategies can be implemented: (1) start with the growth of the core
under conditions which favor the diffusion of adatoms on the
sidewalls and therefore the axial (or longitudinal) growth, then
change the growth conditions in order to enhance the radial growth
and form the shell. For instance, the substrate temperature is
decreased between the core and shell growth steps. (2) use
conditions where a moderate diffusion occurs along the sidewalls,
allowing that both axial and radial growth occur simultaneously,
resulting in cone-shaped (tapered) structures; this can be achieved
using the same, moderately low temperature for the growth of the
whole structure.

In the present study we describe the growth mechanisms of ZnTe NWs
and CdTe-ZnTe structures by molecular beam epitaxy (MBE). The
specific properties of II-VI materials can be fully exploited in
such heterostructured NWs, with applications in photonics,
photodetectors,\cite{Meng09} single photon emitters,\cite{Bounouar}
photovoltaics\cite{PV} including type-II configurations, various
sensors, and magnetic objects.\cite{Wojn12} In the case of
tellurides, (Zn,Mg)Te shells have been shown to enhance the
photoluminescence efficiency by several orders of magnitude with
respect to that of bare ZnTe NW.\cite{Artioli2013} Moreover,
previous studies of heterostructured NWs based on tellurides (of Cd,
Zn, Mg, Mn) suggest that they offer a great flexibility in their
design in order to control the light-hole / heavy-hole character of
the confined ground state of holes.\cite{Wojn12, Artioli2013,
Szymura, Ferrand, Artioli2015, Jeannin}  A low growth temperature
($350^\circ $C) was chosen to minimize the desorption of adatoms,
particularly in view of the insertion of CdTe segments in the ZnTe
NWs. We obtain tapered NWs with smooth sidewalls, and a well
controlled insertion of CdTe and (Cd,Mn)Te. Characteristic
parameters are extracted using the analytical results of a model of
diffusion-limited growth; their dependence on growth temperature and
stoichiometry, and on the shape of the gold nanoparticle, is
discussed.

The paper is organized as follows: the experimental results are
described in section II and used in section III to extract
significant parameters of a diffusion-limited model of growth
adapted to our findings (presented in the appendix). Section IV
discusses the main aspects of the growth and their consequences.

\section{Experimental}
\subsection{Experimental methods and growth conditions}
Our MBE system comprises a Riber 32P II-VI chamber connected under
UHV to a Meca2000 III-V chamber equipped with As and Au effusion
cells. The sample temperatures are measured by a thermocouple in
direct contact with the molybdenum sample-holder. Beam equivalent
pressures (BEP) were converted into flux ratios as described in
Ref.~\onlinecite{Rueda2014} and checked using reflection high-energy
electron diffraction (RHEED) oscillations. An important parameter
for the growth of NWs, is the angle $\alpha$ of each effusion cell
with respect to the normal to the substrate: this angle determines
the ratio of average flux $J_{NW}$ impinging onto the facets of a
normal NW to the flux $J_s$ impinging onto the substrate:
$J_{NW}/J_s=\tan \alpha / \pi $. In our system, the values are $\tan
\alpha=0.21$ for CdTe and Mn, $\tan \alpha=0.48$ for ZnTe and Mg,
$\tan \alpha=0.65$ for Te and $\tan \alpha=0.85$ for Cd and Zn.

Unless otherwise mentioned,  a typical growth sequence consisted in
the deoxidation of the (111)B GaAs substrate at $650^\circ $C under
As flux (1 $\times$ 10$^{-5}$ torr), the growth of a 250-500
nm-thick ZnTe buffer layer at $260^\circ $C under stoichiometric
ZnTe flux, annealing at $420^\circ $C under Te flux, the deposition
of a submonolayer gold layer at room temperature in the III-V
chamber. Gold seeds were formed by dewetting at $350^\circ $C during
5~min in the II-VI chamber. Then the sample rotation was started and
the NW growth initiated immediately at the same temperature.
Monitoring each step by RHEED allows us to check the characteristic
features of a good quality growth: $2\times2$ and $\sqrt{19}\times
\sqrt{19}$ reconstruction of the GaAs surface, \cite{Yang} c(8x4)
and $2\sqrt{3}\times2\sqrt{3}$ R$30^{\circ}$ surface reconstruction
of smoothed ZnTe layers,\cite{RuedaDewett, RuedaPhD} and formation
of a Au-induced $3\sqrt{3}\times3\sqrt{3}$ R$30^{\circ}$ surface
reconstruction.\cite{RuedaDewett, RuedaPhD}

NWs deposited on a holey carbon-coated copper grid were imaged by
energy dispersive x-ray (EDX) spectrometry using a FEI Tecnai Osiris
scanning transmission electron microscope equipped with four silicon
157 drift detectors, as described in Ref.~\onlinecite{RuedaEDX}.
Other NWs, still standing on the substrate, were imaged by scanning
electron microscopy (SEM) using a Zeiss U55 operated at 10~keV, with
a spatial resolution around 1~nm. Selected NWs, attached to the
substrate in the cleaved sample geometry,\cite{MdH} were imaged by
scanning transmission electron microscopy (STEM) on a probe
corrected FEI Titan working at 300 kV, and the lattice spacing in
the high resolution images was analyzed by the geometrical phase
analysis (GPA) \cite{hytchGPA}. Scan noise was removed from the
images by using the unperturbed ZnTe lattice as a
reference.\cite{rouviereGPA}

\subsection{Experimental results}

\begin{figure}
\centering
\includegraphics [width=\columnwidth] {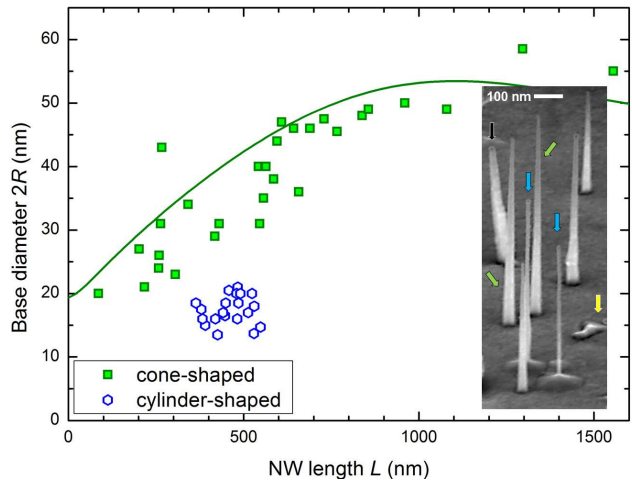}
\caption{SEM data on the diameter at the base, \emph{vs}. the
length, for the NWs of a sample grown at $350^\circ $C with
stoichiometric Zn + Te flux. Green squares mark cone-shaped
(tapered) zinc-blende NWs, blue hexagons cylinder-shaped wurzite NWs
(corresponding arrows in the 60 deg. SEM image in the inset, which
also feature an inclined NW - black arrow - and a creeping object -
yellow arrow); the solid line is calculated as described in section
\ref{fit}. The length plotted here is the length visible above the
2D layer, of thickness $h\simeq$200~nm in this sample. \label{Fig1}}
\end{figure}

Two types of ZnTe NW can be grown: cone-shaped (tapered) NWs, and
cylinder-shaped NWs, as indicated by the green and blue arrows,
respectively, in the inset of Fig.~\ref{Fig1}. The main
characteristics\cite{Rueda2014} of cone-shaped NWs are the
zinc-blende structure with the NW axis parallel to the  $\langle 111
\rangle$ direction, a significant lateral growth on the sidewalls
(formed by $\langle 112 \rangle$ facets visible for thick enough
NWs) and a crater surrounding the base of the NW. In contrast,
cylinder-shaped NWs exhibit the wurtzite structure with the NW axis
along the c-axis, no contribution from lateral growth detected and a
pyramid at the base of the NW clearly visible in the inset of
Fig.~\ref{Fig1}. The flux on the substrate far away from any NW
results in the growth of a two-dimensional layer: its thickness $h$
was determined from the SEM observation of a cleaved edge.

As shown in the plot of Fig.~\ref{Fig1}, the wurtzite NWs are
characterized by a reasonably compact distribution of their length,
and of the base diameter (which actually fairly matches the typical
gold particle diameter). This is in sharp contrast with the
statistics on the zinc-blende NWs: (1) except for the shorter NWs,
the base diameter increases significantly; (2) we observe a broad
distribution of their length, with one order of magnitude between
the longest and the shortest NWs; (3) there is a clear trend
suggesting a correlation between the base diameter and the NW
length.

Such a large dispersion cannot be explained by the distribution of
gold particle size (we did not observe any significant difference in
the external diameter of the gold particle present at the tip of
different NWs), or by the molecular flux. In the following we will
(1) use CdTe markers to reveal the presence of an incubation time,
(2) analyse the distribution obtained when growing at higher
temperature to reveal the role of the shape and size of the gold
nanoparticle, and (3) use a model adapted to the present growth
conditions, which is amenable to analytical expressions, to examine
the role of adatom diffusion.

\subsubsection{CdTe markers}
\begin{figure}
\centering
\includegraphics [width=0.7\columnwidth] {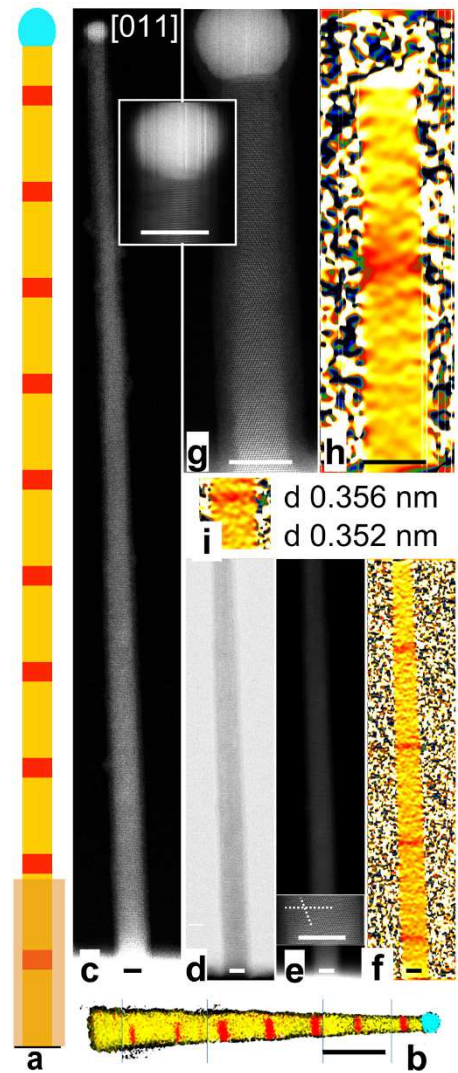}
\caption{(a) Schematic of the intended NW structure with 10
identical sequences ZnTe:CdTe, and a twice shorter ZnTe cap, grown
at 350$^{\circ}$C under stoichiometric fluxes. The 2D layer is
transparently superimposed on the NW showing the part of the NW that
is buried. (b) EDX image of a NW deposited on a holey carbon grid
(composite of 5 higher resolution images). Colored areas are those
where the Cd (red), Zn (yellow), Au (cyan) or O (black) signal
exceeds an arbitrary threshold. (c) HAADF image of a complete NW
(NW1) with the ZB structure, with the substrate at the bottom (white
area). The inset shows the shape of the catalyst particle.(d) BF and
(e) HAADF image of the bottom part of the same NW. The inset in (e)
shows a zoom of the image revealing lattice planes: the NW is
observed along the [$110$] direction and two $\langle 111 \rangle$
planes are marked by dotted lines. (f) GPA analysis of image (d)
showing four small regions with a height of around 4 nm with larger
lattice spacing (in red). (g) HAADF STEM image and (h) GPA analysis
of a second NW on the same sample (NW2; this is the same NW as in
Fig.~\ref{Fig4}b) that also has the ZB structure but is much
shorter. The GPA analysis shows only one region with larger lattice
spacing. (i) Color coding of the lattice spacing in the GPA
analysis. All scale bars are 10 nm for (c)-(h), 100 nm for (b).
\label{Fig2}}
\end{figure}

In order to obtain precise information on a single NW, we used a
marker technique:\cite{marker1, marker2} we inserted 10 thin CdTe
markers (stoichiometric Cd+Te flux from the CdTe cell) during the
growth of a ZnTe NW sample (stoichiometric Zn+Te flux from the ZnTe
cell), see Fig.~\ref{Fig2}a. The EDX image of a NW deposited on a
grid (Fig.~\ref{Fig2}b) clearly reveals 7 regularly spaced CdTe
insertions (and another one at the broken end). On such a broken NW,
we cannot exclude that the two missing insertions were lost in the
harvesting process.

In order to rule out any effect of harvesting, as-grown NWs from the
same sample, still on the substrate, have been visualized by TEM,
and two cone-shaped NWs very different in length (520 and 70 nm
respectively) have been analyzed by GPA. The zinc-blende structure
was confirmed for both NWs.

\begin{figure} [h!]
 \centering
 \includegraphics[width=\columnwidth]{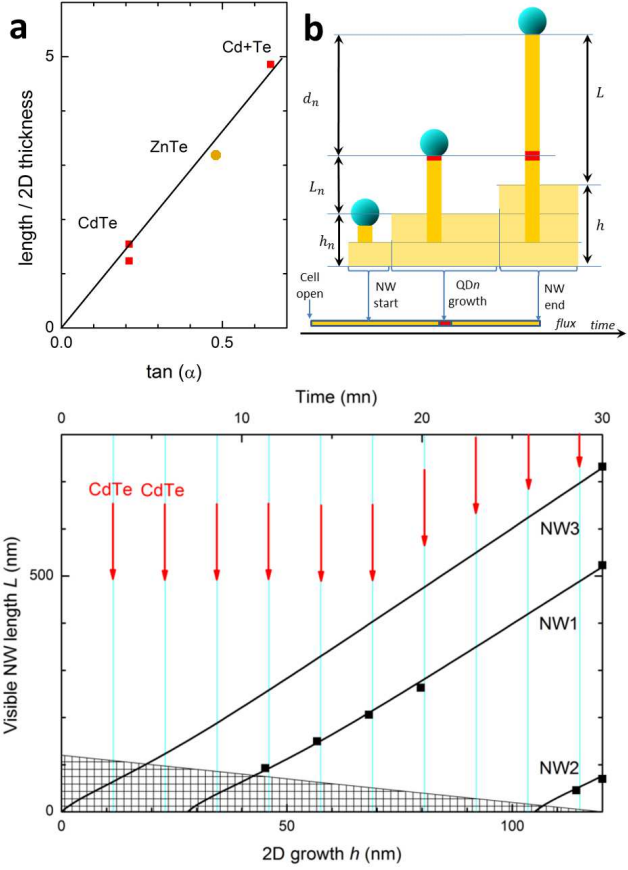}
\caption[]{(a) Effect of the flux incidence angle: length of
individual components of structured NWs (from Fig.~\ref{Fig2} and
Ref.~\onlinecite{RuedaEDX}), divided by the 2D thickness, as a
function of the angle factor $\tan \alpha$, where $\alpha$ is the
angle between the molecular beam and the NW axis. (b) Steps of the
growth of a NW with a CdTe insertion. (c) Plot of the visible length
of NW1 and NW2 of Fig.~\ref{Fig2}, as a function of the 2D layer
thickness; we plot (squares) $L_n$ \emph{vs.} $h_n$ (as defined in
(b), and marked here by the red arrows) at the time of insertion of
CdTe, and $L$ at the end of the growth, \emph{vs}. $h$. NW3 is the
longest NW observed by SEM on the same sample. The hatched triangle
represents the masking by the 2D layer (\emph{i.e.}, the thickness
$(h-h_n)$ remaining to grow after the formation of the CdTe
insertion). The solid lines are the fit with Eq.~\ref{LH}, with the
same values of the diffusion length $\lambda_s=18$~nm on the
substrate and $\lambda_{NW}=78$~nm on the sidewall facets, and
different incubation times. We assumed that the gold nanoparticle
was close to a full sphere from the beginning of the growth, with
the contact diameter 7.7~nm (NW1) and 7.9~nm (NW2), as measured on
the HAADF image, and assuming 7.3~nm for NW3. An additional
constraint (not shown) is that the diameter of the crystalline core
at the base of the NW agrees with Eq.~\ref{RL}. With these values of
the parameters, we deduce an incubation time for NW1 and NW2, and an
immediate start for NW3. }
  \label{Fig3}
\end{figure}

The topmost half of the longer NW (NW1 in Fig.~\ref{Fig2}c-f) could
not be analyzed due to vibrations and charging effects. In the lower
half, four CdTe insertions are clearly identified
(Fig.~\ref{Fig2}f). The period is 68~nm, with 64~nm for the ZnTe
spacers and 4~nm for the CdTe insertions: this ratio is in good
agreement with the ratio of integrated lateral flux, which was
ZnTe:CdTe$\sim$15, suggesting that the growth of this part of the NW
is mainly governed by the adatoms resulting from the flux to the
sidewalls and diffusing to the gold nanoparticle. This preponderant
role of the lateral flux is confirmed by the plot of the length of
CdTe and ZnTe components of various heterostructured NWs grown using
molecular beams with very different angles of incidence
(Fig.~\ref{Fig3}a).

On the shorter NW (NW2 in Fig.~\ref{Fig2}g,h), only one CdTe
insertion is identified, suggesting a long incubation time.

The visible length of the nanowire, $L$, the radius of the NW, the
radius of the gold nanoparticle, and the radius of the gold-tip
contact are available from the HAADF images in Fig.~\ref{Fig2} (see
also Fig.~\ref{Fig4}); the distance of several CdTe insertions from
the tip, $d_n$, are measured on the GPA images. Figure \ref{Fig3}b
describes the growth of a NW with a CdTe insertion, with the
characteristic lengths involved. The length $L$ is the visible
length of the NW, a part of the NW being hidden by the parasitic
growth of the 2D layer. The visible length $L_n$ at the moment of
the growth of the CdTe insertion is obtained as $L_n=L+h-d_n-h_n$,
where the total thickness of the parasitic 2D layer, $h\sim120~$nm,
was measured on a cleavage plane, and its thickness at the moment of
the growth of the CdTe insertion, $h_n$, is deduced from the growth
sequence.

Fig.~\ref{Fig3}c plots $L(h)$ and $L_n(h_n)$ for the two NWs of
Fig.~\ref{Fig2}. Even without the lines which show the fit described
in section \ref{model} below, it appears that the NWs started to
grow after a significant incubation time. This will be addressed
more quantitatively in the remainder of the paper, and possible
reasons will be discussed in Section \ref{Discuss}. Note that the
existence of this incubation time was confirmed by exposing a series
of samples, prepared under the same conditions, to the Zn+Te flux
for 2, 10 or 30 min.: the number of nanoparticles having catalyzed
the growth of NWs was found to increase. Full details are given in
Ref.\onlinecite{RuedaPhD}.

\subsubsection{Gold particle}

\begin{figure}
\centering
\includegraphics [width=1\columnwidth] {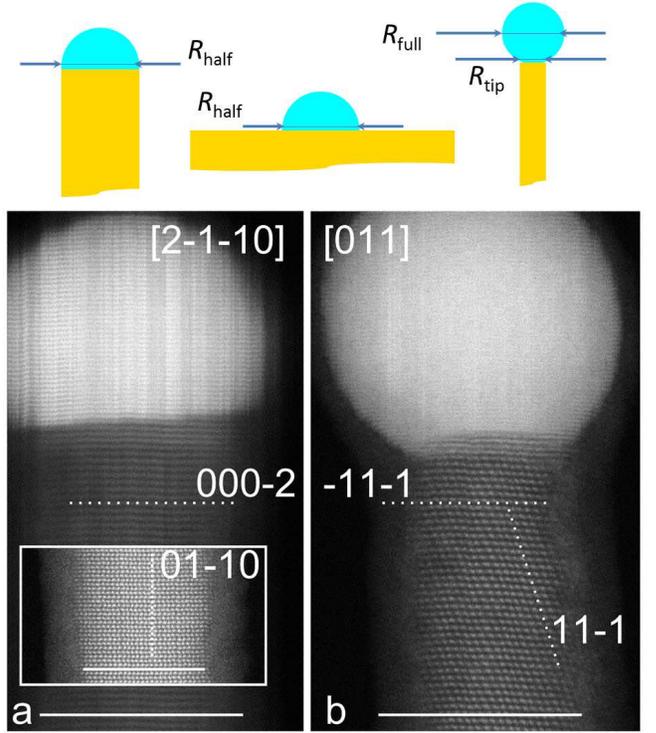}
\caption{HAADF STEM images of gold nanoparticles at the tip of NWs
(a) closer to half-sphere at the tip of a wurzite NW. The image was
made along the [$2\bar{1}\bar{1}0$] direction of observation and the
planes are indexed. The inset shows the wurtzite structure of the NW
that can be observed closer to the base of the NW. (b) The gold
catalyst particle is almost a full sphere at the tip of a
zinc-blende NW. The direction of observation is [011] and both the
$\langle 111 \rangle$ planes are indexed. All scalebars are 10 nm.
The vertical lines most visible in the bright gold catalyst particle
are due to scan errors. \label{Fig4}}
\end{figure}

Figure \ref{Fig4}a shows the gold catalyst on a NW with the wurtzite
crystal structure. The NW was long enough that it vibrates under the
electron beam near the tip, so that only (0002) planes are visible
in the NW; yet the wurtzite structure was clearly observed closer to
the NW base, as shown in the inset of Fig.~\ref{Fig4}a. The NW
exhibits a flat tip, and the shape of the gold particle is close to
a half sphere with a diameter matching the NW diameter. This is
close to the shape assumed by the gold particles sitting at the
surface of the buffer layer before opening the shutter to start the
growth of the NWs.\cite{RuedaDewett, RuedaPhD} This configuration is
well known for III-V NWs with the wurtzite structure, and it is
associated to a nucleation at the edge of the $<0001>$ facet forming
the top of the NW.\cite{Krogstrup} An extensive, statistically
significant study of the interface is however out of the scope of
the present paper.

All tapered nanowires which have been observed by TEM have the
zinc-blende structure. Long tapered NWs show a gold particle with a
shape closer to a full sphere, as shown in Fig.~\ref{Fig4}b. This is
expected if the nucleation takes place on an inclined facet.
\cite{Wen} Indeed, the interface is far from being flat, and a
meniscus (or incompletely formed facet) is clearly apparent.
Hemispherical gold nanoparticles are also observed by SEM at the tip
of tapered NWs, particularly at the tip of the shorter ones. Note
that a change of shape - and of facet morphology - has been observed
on III-V NWs, and may even take place regularly during the growth of
the NW. \cite{Wen} In addition, several facets (3 in the simplest
case) may form, not necessarily at the same time. For such
full-sphere nanoparticles, three characteristic diameters may be
considered: the measured diameter of the full sphere, $2R_{full}$,
the diameter of the initial particle which is approximately
half-sphere,\cite{RuedaDewett, RuedaPhD} $2R_{half}$, and the
diameter of the contact area between the particle and the NW tip,
$2R_{tip}$. If the shape change from half sphere to full sphere
takes place at constant volume, the droplet diameter is slightly
reduced, from $2R_{half}$ to $2R_{full}=2R_{half}/2^{1/3}$,
\emph{i.e.}, by 25\%. More significant is the fact that the diameter
at the interface between the gold particle and the NW - the contact
diameter $2R_{tip}$ - is definitely smaller. It is this diameter
which defines the diameter of the growing NW (hence the lateral size
of a quantum dot embedded in the NW, but also partly the growth
rate, as recalled later on). Typical values of the diameter of the
gold particle are 20~nm. The diameter of the NW neck below the
spherical gold particle is more difficult to measure; typical values
measured by SEM are around 15-20 nm but this includes an amorphous
or ill-crystallized surface layer, which is due either to some
oxidation of the NW, or to post-growth deposition; the crystalline
core imaged by TEM is thinner, 6 to 10 nm (see Fig~\ref{Fig4}, and
also Fig.~5 of Ref.~\onlinecite{Rueda2014}); such a value fairly
matches the diameter of CdTe dots imaged by EDX in
Ref.~\onlinecite{RuedaEDX}.

\subsubsection{The role of the gold particle shape and size}

\begin{figure*}
\centering
\includegraphics [width=2\columnwidth] {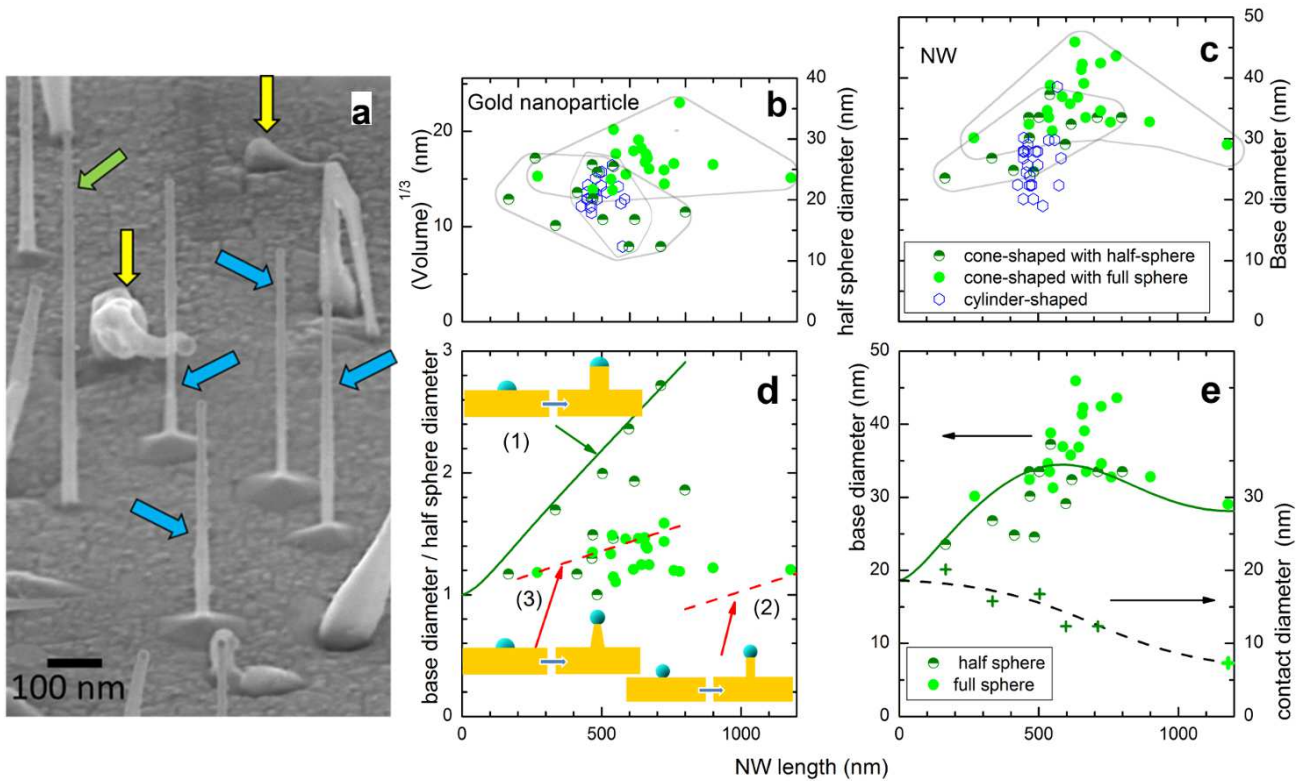}
\caption{(a) 60 deg. SEM image of a sample grown at 375$^{\circ}$C
under Te-excess (yellow arrows mark 3D objects, green arrows
zinc-blende NWs and blue arrows wurtzite NWs); (b) size of the gold
nanoparticle, and (c) plot of the base diameter, \emph{vs}. NW
length (with identification of the cylinder-shaped wurtzite NWs,
open hexagons, and of the cone-shaped zinc-blende NWs with a
half-sphere or full-sphere gold particle at the tip, as indicated by
the half-filled or full circles, respectively); the grey lines are
guides for the eye; (d) Base diameter to nanoparticle size ratio,
and (e) base diameter, \emph{vs} NW length; crosses in the lower
part of (e) display the nanoparticle diameter for the NWs at the
edge of the distribution in (d), \emph{i.e.}, the 5 NWs on the solid
line at the left top side of the distribution, and the longest one;
lines in (d) and (e) are calculated as described in section
\ref{model}. \label{Fig5}}
\end{figure*}

With the previous samples, we have explored the growth at a
temperature ($T_{growth}=350^{\circ}$C) where the distribution of
gold nanoparticles is relatively narrow. Figure \ref{Fig5}a shows a
SEM image of a sample obtained by first dewetting the gold
nanoparticles at 350$^{\circ}$C, and rapidly increasing the
temperature to 375$^{\circ}$C to start the growth with an additional
Te flux. It displays cylinder-shaped NWs (blue arrows), and
cone-shaped NWs (green arrows) - quite similar to samples grown
under stoichiometric conditions at $350^{\circ}$C. This indicates
that the increase of the sample temperature and the presence of a
higher Te flux compensate each other, a hint that the evaporation of
Te atoms is probably the limiting mechanism. Also visible are
aborted or badly formed NWs (yellow arrows) and thick, inclined NWs
(bottom right corner).

Fig.~\ref{Fig5}b plots the volume of the gold nanoparticle,
\emph{vs} the NW length. As observed previously in Fig.~\ref{Fig4},
the nanoparticles assume two different shapes (close to half-sphere
and close to full sphere), which are identified in Fig.~\ref{Fig5}
by different symbols. Again one observes a stronger dispersion in
the NW length for the cone-shaped NWs than for the cylinder-shaped
ones. But also one observes (1) a significant dispersion in the
nanoparticle size, which is due to an evolution of the nanoparticles
during the ramp to the higher growth temperature, and (2) a
correlation between the shape of the gold particle at the tip
(half-sphere or full-sphere) and the length of the cone-shaped NWs:
the two sets of NWs occupy two different areas, delineated by the
grey lines.

As a result of the dispersion in the nanoparticle size, the trend
between the base diameter and the NW length (Fig.~\ref{Fig5}c)
appears as partially blurred when compared to Fig.~\ref{Fig1}. To
cope with this dispersion, in Fig.~\ref{Fig5}d we plot the ratio of
the base diameter to the size of the gold nanoparticle, $R/R_{half}$
where $R_{half}$ is the radius of the half-sphere with the same
volume (see Fig.~\ref{Fig5}b). This plot will be used in the
following section, in the frame of a growth mechanism taking into
account the diffusion of adatoms and the incubation time mentioned
previously, but also the effect of the nanoparticle size and shape:
the results of this model are summarized by the lines in
Fig.~\ref{Fig5}d,e, to be discussed in section \ref{model}.

\section{Diffusion-limited model} \label{model}
The growth of nanowires is strongly influenced by two parameters:
the nucleation at the Au catalyst particle, and the diffusion of
species on the surfaces. On one hand, complex models based on growth
kinetics, involving the interaction of atoms in various
thermodynamic states at the Au-NW interface, have been proposed for
the VLS growth of III-V NWs.\cite{Glas2010} These models rely on
transition state kinetics driven by the minimization of the free
energy of the total system. On the other hand, diffusion-dependent
nanowire growth models have also been developed
.\cite{Dubrovskii2006} In these type of models, the morphology
(length and shape) of the nanowires is directly related to different
values of the diffusion length on the substrate and along the NW
sidewalls. These models are also known to apply to a wide range of
growth conditions of III-V and Si NWs\cite{Dubrovskii2009} and even
to CdTe NWs.\cite{Dubrovskii2012}

In this section we focus on the role of adatom diffusion on the
growth of ZnTe NWs by MBE at low temperature: any possible effect of
the nucleation at the gold particle / NW tip is ignored. We obtain
analytical expressions which we can readily exploit to extract
quantitative parameters (diffusion lengths, incubation times) from
the plots in Fig.~\ref{Fig1}, \ref{Fig3}, \ref{Fig5}.

\subsection{Axial and lateral growth rates in the diffusion-limited model} \label{ModelResults}
 The description is made simpler if we keep in mind that:

\begin{itemize}
  \item The density of Au droplets (and even more the density of NWs) on the substrate is low enough that we can consider a single, isolated NW.
  \item The NW is perpendicular to the substrate surface.
  \item The temperature is low enough that re-evaporation is negligible: the diffusion length is defined by the incorporation of adatoms on the substrate or on the NW facets.
  \item We observe a NW growth rate significantly larger than that of the 2D layer;
  hence we calculate only the contribution of adatoms diffusing from
  the substrate and the NW sidewalls to the gold particle. Indeed, the
  contribution of the direct flux onto the Au droplet is not too different from the contribution to the 2D layer away from the NW (they are equal for a half-sphere droplet and complete sticking); hence the length we calculate is close to the length visible above the 2D layer. \\
\end{itemize}

The diffusion-based NW growth model adapted to these conditions is
described in Appendix \ref{calculation}. It involves the general
solution of the diffusion equations of adatoms on the NW sidewalls
in 1D, and on the substrate surface in 2D with a circular symmetry.
It assumes complete trapping at the tip of the NW (catalyst acting
as a perfect sink).\cite{Dubrovskii2006} It results in simple
analytical relationships between the length $L$ of the NW (axial
growth) and the thickness $h$ of the 2D layer far from the NW
(growth in the absence of catalyst), on one hand, and between $L$
and the radius $R$ at the NW basis (lateral growth), on the other
hand. These relationships are derived from Eq.~\ref{LH} and \ref{RL}
in Appendix \ref{calculation} where we set all sticking coefficients
to unity (this is a reasonable approximation for the ZnTe surfaces
at $350^\circ $C):
 \begin{equation}\label{LH1}
\frac{h}{R_0}=\frac{1}{2}\frac{\pi}{\tan \alpha} \ln
\frac{1+\frac{\tan
\alpha}{\pi}\left(\frac{\lambda_{NW}}{\lambda_s}\right)^2\left[\cosh(\frac{L+\tilde{\lambda}_s}{\lambda_{NW}})-1\right]}
{1+\frac{\tan
\alpha}{\pi}\left(\frac{\lambda_{NW}}{\lambda_s}\right)^2\left[\cosh(\frac{\tilde{\lambda}_s}{\lambda_{NW}})-1\right]}
\end{equation}
and
\begin{eqnarray}\label{RL1}
\frac{R-R(0)}{R_0}=\frac{1}{2}
\frac{\pi}{\tan \alpha}\left(\frac{\lambda_s}{\lambda_{NW}}\right)^2\times\nonumber\\
\ln \frac{1+\frac{\tan
\alpha}{\pi}\left(\frac{\lambda_{NW}}{\lambda_s}\right)^2\left[\cosh(\frac{L+\tilde{\lambda}_s}{\lambda_{NW}})-1\right]}
{1+\frac{\tan
\alpha}{\pi}\left(\frac{\lambda_{NW}}{\lambda_s}\right)^2\left[\cosh(\frac{\tilde{\lambda}_s}{\lambda_{NW}})-1\right]}
\end{eqnarray}
where $\lambda_s$ and $\lambda_{NW}$ are the diffusion length on the
substrate and along the NW sidewalls, respectively, $R_0$ is the
radius of the contact area between the gold catalyst and the NW tip,
$R(0)$ is the initial radius of the gold nanoparticle. We have
replaced the ratio of lateral to substrate flux by its value,
$\frac{\tan \alpha}{\pi}$, and we have defined an effective
diffusion length $\tilde{\lambda}_s$, which is small with respect to
$\lambda_{NW}$ and kept constant in the calculation, as discussed in
Appendix \ref{calculation}.

Examples of Eq.~\ref{LH1} are displayed in Fig.~\ref{FigA2}a: the
initial growth rate (for $L\ll\lambda_{NW}$) is governed by the
diffusion on the substrate; for most of the NWs described here, the
final length $L\gg\lambda_{NW}$ so that the growth rate during most
of the growth is the asymptotic growth rate due to the lateral flux,
given by $\frac{dL}{dh}\approx \frac{2 \tan \alpha}{\pi}
\frac{\lambda_{NW}}{R_0}$.

The lateral growth at the basis of the NW (Eq.~\ref{RL1}) reflects
the fact that the adatom density on the sidewalls at the basis of
the NW is linked to the adatom density on the substrate, so that the
difference in the incorporation rates leads to
$\frac{dR}{dh}\simeq\left(\frac{\lambda_s}{\lambda_{NW}}\right)^2$.
Figure \ref{FigA2}b shows that this holds for most of the growth,
with a small correction due to the lateral flux, and except for the
initial step where both the NW length and the lateral growth are
small. We will use Eq.~\ref{RL1} in the analysis of experimental
data, but the asymptotic form suggests that the diameter of the NW
at its basis is a rather good measure of the time between the actual
start of the NW growth and its end.

\subsection {The model applied to experimental data} \label{fit}

\subsubsection {The markers technique}

The most complete data are those obtained on single NWs with CdTe
insertions, Fig.~\ref{Fig3}. A reasonable fit (black solid lines) is
obtained with the same values of the diffusion lengths for both NWs
($\lambda_s$=18~nm and $\lambda_{NW}$=78~nm), assuming that the
growth of the NW does not start immediately after the opening of the
cell. A further criterion for the fit (not shown) is that the
measured and calculated values of the base diameter coincide. The
shape of the gold particle at the tip of NW1 and NW2 is close to a
full sphere: the fit in Fig.~\ref{Fig3} assumes that the
nanoparticles turned very rapidly to a full-sphere; hence in
Eq.~\ref{RL} the initial diameter, $2R(0)$, and the contact diameter
during the growth, $2R_0$, are both taken equal to $2R_{tip}$.
Moreover, we have assigned to $2R_{tip}$ the diameter of the
crystalline core of the NWs measured by TEM, $\simeq7.7$nm for NW1
and 7.9 nm for NW2. The length measured for NW3 is well fitted
assuming the same values of diffusion lengths, and $2R_{tip}=7.3$
nm. Actually, there is a 2.5~nm thick amorphous layer around the NW,
which can be due to the oxidation of the NW or to a post-growth
deposition. If we take the external diameter instead of that of the
crystalline core in Eq.~\ref{RL}, an as-good fit is obtained with
$\lambda_{NW}$ unchanged and $\lambda_s$ increased by 15 to 20\%
(not shown).

The main conclusion here is that the broad range of length values
observed involves a significant incubation time: the 2D layer
thickness was already 30 nm when NW1 started, and 100 nm for NW2.
This delayed growth is the main reason for the length dispersion of
the 3 NWs shown in Fig.~\ref{Fig3}.

It will be further confirmed now using a complementary approach:
indeed a similar information can be deduced from the observation of
an ensemble of NWs (without markers) and the statistics on the
length / base diameter relationship.

\subsubsection {Axial vs. radial growth}

We start with the results shown in Fig.~\ref{Fig5}. Keeping in mind
the two possible shapes of the gold nanoparticle, close to
half-sphere or close to full sphere, we may expect three typical
behaviors: NWs driven by a nanoparticle staying half-sphere up to
the end of growth (case 1 below), NWs driven by a nanoparticle
turning to full sphere from the beginning (case 2), and nanoparticle
turning from half-sphere to full sphere shortly after the growth
start (case 3). As the growth rate is expected to increase as the
contact diameter decreases, the full-sphere-NWs are expected to grow
faster than the half-sphere-NWs in spite of a larger nanoparticle
volume: the relevant diameter is $2R_{tip}\sim$6 to 8~nm in the
first case, against $2R_{half}\sim$15 to 18~nm in the second case.

\begin{itemize}
  \item (1) One clearly identifies in Fig.~\ref{Fig5}d a series of
half-sphere NWs with a large base diameter and small length, which
we tentatively ascribe to the extreme case of full-time half-sphere
NWs. Assuming $2R(0)=2R_0=2R_{half}$ in Eq.~\ref{RL}, with the
additional assumption that the longest NW of the series started to
grow immediately, a reasonable fit is obtained as shown by the solid
line, with $\lambda_s$=38~nm and $\lambda_{NW}$=126~nm. The
different values of the NW length, from the shortest to the longest,
correspond to (\emph{i}) increasing values of the growth duration,
with the shortest NW starting on a 2D layer of thickness $\sim 75$
nm (40 nm still to go) up to the full $h=115$~nm for the longest one
(immediate start), (\emph{ii}) a decreasing nanoparticle size
measured within this series of half-sphere NWs (green crosses in
Fig.~\ref{Fig5}e). Taking both effects into account, and using a
purely heuristic dependence between the contact area and the final
length (dashed line in Fig.~\ref{Fig5}e), we obtain the green solid
line in Fig.~\ref{Fig5}e.
  \item (2) For the opposite extreme, the case of a full sphere since the
beginning, we keep the same values of the diffusion lengths and
decrease the value of $2R(0)=2R_0=2R_{tip}$ to 7.3~nm to obtain the
largest length observed: we then obtain the lowest red dashed line
in Fig.~\ref{Fig5}d; note that the previous dashed and solid lines
in Fig.~\ref{Fig5}e satisfactorily account for this experimental
point.
  \item (3) The second red dashed line in Fig.~\ref{Fig5}d is obtained using $2R(0)=2R_{half}$ and
$2R_0=2R_{tip}$=7.3~nm in Eq.~\ref{RL}, \emph{i.e}., starting the
growth with a half-sphere and rapidly turning to a full sphere.
\end{itemize}
Intermediate data - in particular half-sphere data below the full
line, or full-sphere data between the two dashed lines - can be
ascribed to NWs with a non-circular shape of the nanoparticle, or a
shape which evolves during the growth. Finally, the green solid line
in Fig.~\ref{Fig5}e separates the half-sphere NWs from the
full-sphere NWs, it also constitutes a good average fit of the
base-diameter / length dependence.

A good overall fit is also obtained using the same procedure for the
other sets of data. For instance, in Fig.~\ref{Fig1}, we had to
assume a diameter decreasing from $2R_{half}$=20nm to $2R_{tip}$=6
nm, and diffusion lengths $\lambda_{NW}$=78 nm and $\lambda_s$=23
nm, close to those of the sample in Fig.~\ref{Fig3}, as expected
since the growth conditions are the same (the slight difference is
at least partly explained that we used diameter values obtained
either for the crystalline core observed by TEM or the overall NW
imaged by SEM). For a sample grown under Te excess (not
shown\cite{RuedaPhD}), slightly lower values of the diffusion
lengths, $\lambda_{NW}$=57 nm and $\lambda_s$=18 nm, are obtained,
assuming a similar distribution of the gold nanoparticle size. All
values are gathered in Table I below.
\section{Discussion and conclusion} \label{Discuss}
In summary, the large dispersion in the length of the zinc-blende
ZnTe NWs is ascribed to the presence of two steps in the growth
catalyzed by gold nanoparticles, which initially feature a
half-sphere shape: the start of the NW growth, and the evolution of
shape of the gold nanoparticle from half-sphere to almost full
sphere. The longest NWs have started immediately with a nanoparticle
rapidly turning to full-sphere,  hence a small contact radius, with
facets or a meniscus at the interface. The shortest NWs started with
a delay and with a nanoparticle staying with a half-sphere shape on
top of the 2D layer and the NW tip. Some nanoparticles never give
rise to a NW or even are buried into the 2D layer. The direct effect
of the particle size on the growth rate exists but it is mostly
hidden by the existence of the incubation delay.

Within this scheme, the characteristics of the NWs are well
accounted for using a simple diffusion-limited model, with two
diffusion lengths, $\lambda_{NW}$ on the NW sidewalls and $\lambda_s
$ on the substrate. Table \ref{Tab1} shows the values used to fit
our data: the decrease induced by adding a Te flux suggests that the
limiting process under these conditions is the diffusion of Zn (the
diffusion length of Zn is reduced if the density of Te adatoms
increases). Increasing the temperature increases the diffusion
length under Te excess, presumably due to an activation of Te
desorption.

\begin{table} [h]
\begin{center}
    \begin{tabular}{| l | l | l | l | }
    \hline
    Growth & $\lambda_{NW}$ (nm) & $\lambda_s $ (nm) &  Fig.\\\hline
    $350^\circ$C-stoichio& 78& 18&\ref{Fig2}, \ref{Fig3} \\ \hline
    $350^\circ$C-stoichio& 78& 23&\ref{Fig1} \\ \hline
    $350^\circ$C-Te& 57& 18& Ref.~\onlinecite{RuedaPhD} \\ \hline
    $375^\circ$C-Te&126 & 38& \ref{Fig5} \\
    \hline
    \end{tabular}
\end{center}
\caption{Diffusion length values used in the fits for different
growth conditions. } \label{Tab1}
\end{table}

With such values of the diffusion length $\lambda_{NW}$ on the NW
sidewalls, the growth rate at the tip of a NW with a length of
several 100~nm is mainly due to the lateral flux, $J_{NW}$, and
$\frac{dL}{dh}\simeq\frac{2\tan\alpha}{\pi}\frac{\lambda_{NW}}{R_0}$.
This is important as it determines the length of a quantum dot
inserted in the NW, but also its composition if the material is an
alloy, for instance a diluted magnetic semiconductor. The diameter
of the quantum dot is determined by the contact diameter ($2R_0$).

Our uncertainty on the value of $R_0$ (crystalline core or the whole
NW including the amorphous shell), and that on the value of $h$,
have a straightforward impact on the value of $\lambda_{NW}$. In
addition, sticking coefficients lower than unity will change the
value of $\lambda_{NW}$, by a factor $\frac{\alpha_s}{\alpha_{NW}}$
(see Eq.~\ref{P3}).

The value of $\lambda_s$, the diffusion length on the substrate, is
mainly deduced from the measure of the lateral growth at the basis
of the NW. Figure \ref{FigA2} suggests that for the present values
of the parameters, both the effect of the depletion of adatoms
around the NW basis and the contribution of the lateral flux are
small, and moreover they partially compensate each other. Here
again, sticking coefficients  lower than unity will affect the
absolute value of $\lambda_s$, as we actually determine
$(\frac{\lambda_s}{\lambda_{NW}})^2 \frac{\alpha_{NW}}{\alpha_s}$.
Note that our values of $\lambda_s$ are quite small, but they are of
the same order as the size of the structures observed at the bottom
of the NWs: a crater or a zone with no growth in the case of
zinc-blende NWs, signaling a transfer of material from the substrate
to the NW, and a pyramid progressively formed at the bottom of
wurtzite NWs, as the signature of a transfer from the NW to the
substrate.

We have noted that the effect of the additional Te flux and of the
growth temperature (Table \ref{Tab1}) suggests that it is the
diffusion of Zn which limits the growth rate - even if a role of Te
diffusion is not totally ruled out. The growth of ZnTe NWs thus
strongly differs from the growth of self-assembled GaAs NWs, which
is characterized by a strong diffusion of Ga towards the tip, and a
growth rate determined by the As$_4$ flux to the Ga droplet
\cite{Colombo}, either directly from the effusion cell, or by
desorption \cite{Ramdani}. In our case, we observe the opposite
behavior, that adding a Te flux decreases the axial growth rate by
decreasing $\lambda_{NW}$.

An important result of the present study is the evidence of an
incubation time between the onset of the molecular beam and the
start of the NW growth. Different systems have been shown previously
to exhibit significant incubation times. One reason may be the
formation of the proper catalyst/semiconductor interface. When GaAs
NWs are grown by gold-catalyzed MBE on a silicon substrate under
vapor-liquid-solid conditions,\cite{Breuer} an initial step is
observed where "traces" creep on the Si surface until they meet a
GaAs nucleus. This is interpreted as due to a large difference in
the interface energy on the Si substrate and on the GaAs nucleus. We
do observe such traces (yellow arrows in Fig.~\ref{Fig1} and
\ref{Fig5}), but in our case the gold nanoparticles are formed on a
ZnTe buffer layer. For other preparation and growth conditions, the
incubation is attributed to the formation of a pinhole in the Si
oxide, and the formation of the Ga droplet \cite{Fontcuberta,
Rudolph}. Another contribution may be brought by the existence of a
threshold in the composition of the catalyst droplet. In the case of
Si NWs on Si substrates, depending on the growth conditions, this
threshold has been discussed in terms of supersaturation of the gold
droplet, \cite{Dhalluin} or in terms of diffusion within the solid
catalyst to achieve its fusion (transition from vapor-solid-solid,
VSS, to vapor-liquid-solid, VLS conditions).\cite{Kalache} Such a
transition from VLS to non-VLS conditions was observed to depend on
the As/Ga ratio for GaAs NWs on Si.\cite{Rudolph} In all cases the
incubation time is expected to increase with the catalyst size,
which is indeed a trend which is suggested by the catalyst size
dependence on final NW length (data in Fig.~\ref{Fig5}, and the
heuristic dependence used in the overall fits). A dispersion in the
value of the diffusion length on the substrate may enhance the
dispersion in the incubation times. A specific study is needed to
disentangle and evaluate the possible contributions, in order to
obtain a reasonably narrow length distribution.

Finally, we may come back to the NW of Fig.~\ref{Fig2}b. From
quantitative profiles (not shown), the gold nanoparticle diameter is
35 nm, the NW diameter below the gold nanoparticle is 26 nm (total)
and 20-22 nm (not counting the oxide shell), the NW diameter at the
base is 55 nm (with the oxide) and 40 nm (without the oxide). 8 CdTe
insertions are detected along the length of 550 nm, regularly spaced
with a period $67 \pm1$ nm. The size of the gold nanoparticle is
clearly larger than average, and such that we expect it to remain as
a 3D object, or to give rise to an inclined NW (see an example with
the black arrow in the inset of Fig.~\ref{Fig1}). In this case a
larger contact area and a stronger lateral flux due to the inclined
configuration tend to increase the tapering. Note also that the
lateral growth of ZnTe on the side of the CdTe insertion is not
symmetrical, as evidenced by the off-axis position of the CdTe
insertions at the bottom of the NW.

To conclude, we have demonstrated two approaches to unravel the
characteristics of the MBE growth of NWs at low temperature, with
the help of analytical results from an adapted diffusion-based
model. Applied to the growth of ZnTe NWs, the main results are:

\begin{itemize}
  \item the existence of an incubation time before the actual start of the
NW, which remains to be understood, and if possible suppressed;
  \item except for very short NWs, the central role of lateral flux
in determining the axial growth rate (hence the length and
composition of an embedded quantum dot); note that the incubation
time does not affect this phase of the growth;
  \item the formation of a narrow contact area, which determines the
diameter of the quantum dot\cite{RuedaEDX} but also contributes to
determine the axial growth rate;
  \item the possibility of lateral growth without changing the growth
temperature, allowing us to fabricate core-shell structures with
good interfaces or tapered NWs; at the basis of the NW, the lateral
growth is governed by the diffusion of substrate adatoms; higher
along the NW, the role of the lateral flux progressively increases.
\end{itemize}

\appendix
\section {Calculation} \label{calculation}

\setcounter{figure}{0}
\renewcommand{\thefigure}{A\arabic{figure}}

Or goal here is not to derive a general theory of NW growth, but to
obtain a simple model, adapted to the present case, with analytical
results, likely to serve as a guide for future studies.

We consider a low density of NWs, perpendicular to the substrate. A
diffusion-based NW growth model is built from the general solution
of the diffusion equations of adatoms on the NW sidewalls in 1D, and
on the substrate surface in 2D  with a circular symmetry. The flux
$J_s$ impinging onto the substrate (actually, the ZnTe buffer layer)
far from the NW induces the presence of adatoms with a density
$n_s^0=J_s \tau_s$, where $\tau_s$ is the residence time of an
adatom which diffuses on the surface over a length
$\lambda_s=\sqrt{D \tau_s}$ before incorporation (or desorption). We
define similarly a uniform-limit adatom density on the NW sidewalls,
$n_{NW}^0=J_{NW} \tau_{NW}$, with $\tau_{NW}$ the residence time and
$\lambda_{NW}=\sqrt{D \tau_{NW}}$ the diffusion length. For the sake
of simplicity, we keep the same diffusion constant $D$. The ratio of
adatom densities is thus (with $\alpha$ incidence angle,
\emph{i.e.}, the angle between the cell axis and the substrate
normal)

\begin{equation}\label{adatomratio}
\frac{n_{NW}^0}{n_s^0}=\frac{J_{NW}}{J_s}\left(\frac{\lambda_{NW}}{\lambda_s}\right)^2=\frac{\tan
\alpha}{\pi}\left(\frac{\lambda_{NW}}{\lambda_s}\right)^2
\end{equation}

The local densities of adatoms (number of adatoms by surface unit),
$n_s(r)$ on the substrate, and $n_{NW}(z)$ on the NW sidewalls, are
solutions to the diffusion equations and can be written as follows:

\begin{eqnarray}\label{M1}
n_{NW}(z)&=&n_{NW}^0+c_{NW}^+ \exp(\frac{z}{\lambda_{NW}})+c_{NW}^-
\exp(-\frac{z}{\lambda_{NW}}) \nonumber\\
 n_s(r)&=&n_s^0+c_s^+
\textrm{I}_0(\frac{r}{\lambda_s})+c_s^-
\textrm{K}_0(-\frac{r}{\lambda_s})
\end{eqnarray}

where $r$ is the radial coordinate centered on the NW axis and
increasing when moving away from the NW, and $z$ is the coordinate
along the NW axis, perpendicular to the substrate and increasing
towards the tip of the NW. $\textrm{I}_0$ and $\textrm{K}_0$ are
modified Bessel functions.

The four coefficients $c_{NW}^+$, $c_{NW}^-$, $c_s^+$, $c_s^-$, are
to be determined by boundary conditions. We take the simplest ones:

\begin{itemize}
  \item For a weak density of NWs, $n_s(\infty)=n_s^0$ hence
$c_s^+=0$.
\item The hopping probabilities of adatoms, from the NW to
substrate and from the substrate to the NW, are assumed to be
symmetric: $n_{NW}(z=0)=n_s(r=R)$, where $R$ is the nanowire radius
at the base.
  \item The adatom current at the vicinity of the NW-substrate interface are in equilibrium (conservation
of the adatom currents): $(dn_{NW}/dz)_{z=0}=-(dn_s/dr)_{r=R}$
  \item Complete trapping at the tip of the NW (catalyst droplet acting as perfect
  sink): $n_{NW}(z=L)=0$, where $L$ is the height of the nanowire.
This last condition is valid for a purely diffusion-driven model
.\cite{Dubrovskii2006} In the more complex case of kinetic-driven
growth \cite{Glas2010} (not treated here), where the thermodynamical
mechanisms in the catalyst droplet limiting the axial growth need to
be taken into account, this condition is modified to a condition on
$(dn_{NW}/dz)_{z=L}$.
\end{itemize}

These boundary conditions allow us to determine the constants in
equations \ref{M1} and to obtain the following expression for the NW
adatom density:
 \begin{equation}\label{ns1}
   n_{NW}(z)=n_s^0 \: \omega_s(z) + n_{NW}^0 \: \omega_{NW}(z) \\
   \end{equation}

with $\omega_s(z) =
\frac{\sinh(\frac{L-z}{\lambda_{NW}})}{\sinh(\frac{L+\tilde{\lambda}_s}{\lambda_{NW}})}$
and

$\omega_{NW}(z) =
1-\frac{\sinh(\frac{z}{\lambda_{NW}})}{\sinh(\frac{L}{\lambda_{NW}})}-\frac{\sinh(\frac{L}{\lambda_{NW}})
+
\frac{\tilde{\lambda}_s}{\lambda_{NW}}}{\sinh(\frac{L+\tilde{\lambda}_s}{\lambda_{NW}})}
\frac{\sinh(\frac{L-z}{\lambda_{NW}})}{\sinh(\frac{L}{\lambda_{NW}})}$

Here we have defined an effective diffusion length
$\tilde{\lambda}_s=-\lambda_s \textrm{K}_0(\frac{R}{\lambda_s}) /
\textrm{K}'_0(\frac{R}{\lambda_s})$, and we have assumed that
$\tilde{\lambda}_s<<\lambda_{NW}$. Note that if the diffusion
equation in the substrate was in 1D instead of in 2D,
$\tilde{\lambda}_s$ would simply be $\lambda_s$. In 2D,
$\tilde{\lambda}_s$ depends on $R$, hence its value changes as
growth proceeds. Indeed, when we calculate the values of
$\tilde{\lambda}_s$ for $R$ from 10 to 35 nm (the range of NW base
radius values found for our ZnTe NWs, see Fig.~\ref{Fig1}), we find
$\tilde{\lambda}_s=0.3 \lambda_s$ to $0.5 \lambda_s$, and for
typical values of $\lambda_s$ (see below), $\tilde{\lambda}_s$
remains of the order of a few 10~nm. In addition,
$\tilde{\lambda}_s$ influences mainly the NW growth when they are
short (\emph{i. e.} when the effect of lateral growth is still
negligible). When the NWs are longer, it is $\tilde{\lambda}_s + L$
that dominates the growth, hence the variations of
$\tilde{\lambda}_s$ are masked by the larger values of $L$. In the
following, we take $\tilde{\lambda}_s$ constant and small with
respect to $\lambda_{NW}$.

\begin{figure}[h]
 \centering
 \includegraphics[width=\columnwidth]{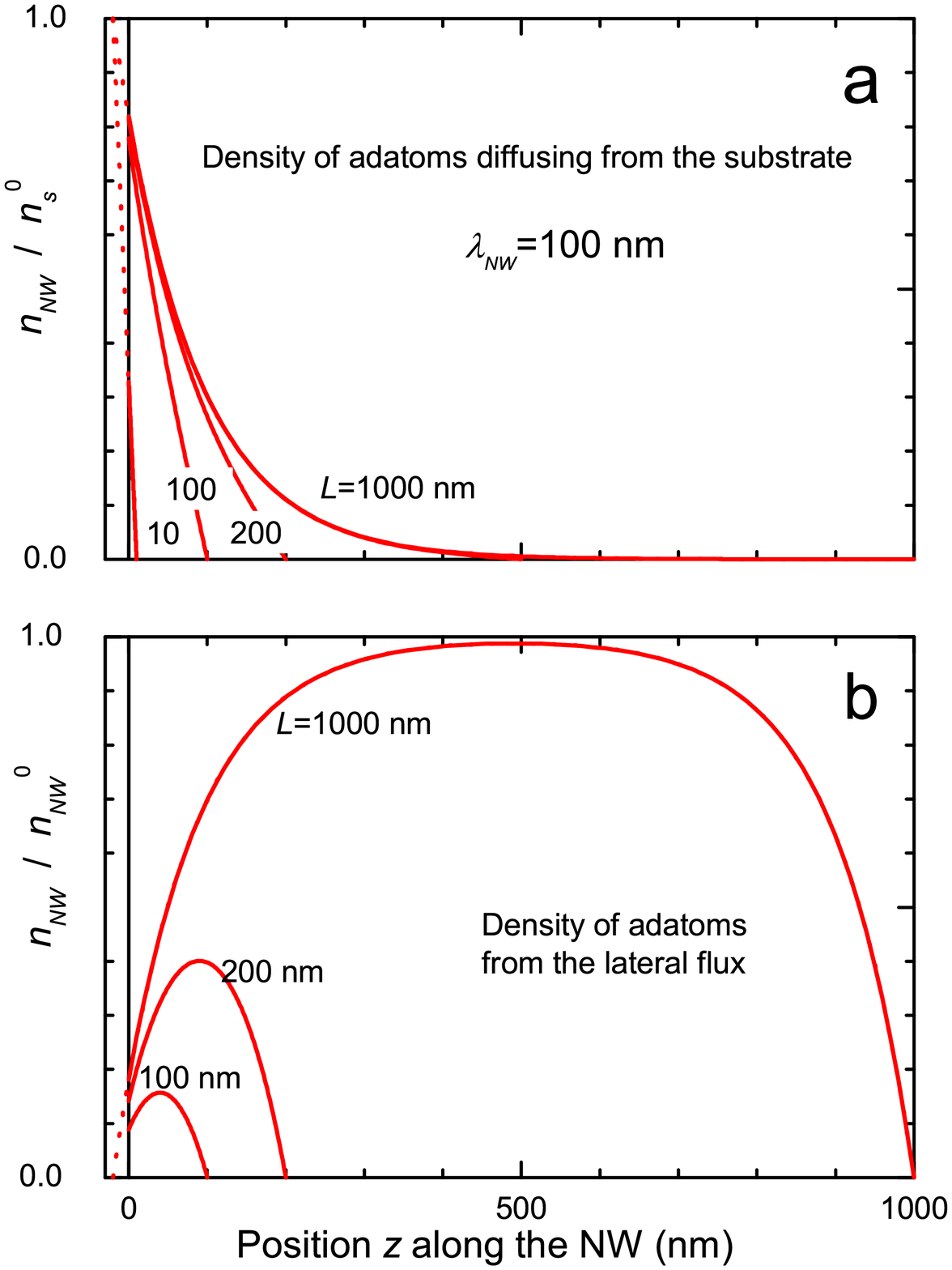}
\caption[]{(a) Contribution to the NW density of adatoms diffusing
from the substrate (Eq.~\ref{ns1}): $\omega_s(z)$ is plotted for
different values of the NW length as specified with $\lambda_s=60$
nm, $\tilde{\lambda}_s= 20 $ nm and $\lambda_{NW}= 100 $ nm. (b)
Contribution to the NW density of adatoms from the lateral flux
(Eq.~\ref{ns1}), $\omega_{NW}(z)$. All curve would coincide at
$z=-\tilde{\lambda}_s$.}
  \label{FigA1}
\end{figure}

The two contributions to Eq.~\ref{ns1}, $\omega_s$ from the flux
arriving to the substrate, and $\omega_{NW}$ from the flux arriving
laterally, are plotted in Fig.~\ref{FigA1} for typical values
$\lambda_s=60$ nm, $\tilde{\lambda}_s= 20 $ nm and $\lambda_{NW}=
100 $ nm. Note that all curves converge at $z=-\tilde{\lambda}_s$ as
shown by the dotted lines, which identifies $\tilde{\lambda}_s$ as
an effective diffusion length on the substrate, as seen from the NW.

The growth rates for the NW length $L$ (axial growth), for the NW
radius at the base $R$ (lateral growth) and  for the 2D
pseudo-substrate with height $h$ are given by:

\begin{eqnarray}\label{P3}
\frac{dL}{dt}&=&-\alpha_{L}\frac{2D\Omega_0}{R_0}\frac{dn_{NW}}{dz}(z=L)\nonumber\\
\frac{dR}{dt}&=&\alpha_{NW}\frac{n_{NW}(z=0)}{\tau_{NW}}\Omega_0\nonumber\\
\frac{dh}{dt}&=&\alpha_s\frac{n_s^0}{\tau_s}\Omega_0=V_s
\end{eqnarray}

In these equations, $\Omega_0$ is the volume occupied by an atom in
the crystal; $\alpha_s$, $\alpha_{NW}$ and $\alpha_L$  are the
incorporation rate at the substrate, at the NW facets and at the tip
of the NW (at the interface with the Au catalyst particle); $V_s$ is
the growth speed on the substrate (at the end of the growth, the
final thickness of this re-grown layer is $h=V_s t$).

From Eq.~\ref{P3} and Fig.~\ref{FigA1}a, we conclude that the
contribution to the growth from adatoms diffusing from the substrate
is: \emph{i}) important for short NWs, $L\ll\lambda_{NW}$, with a
linear decrease of $n_{NW}$ with $z$, and \emph{ii}) weak for long
NWs, $L\gg\lambda_{NW}$, with a exponential decrease of $n_{NW}$
with $z$.

Figure \ref{FigA1}b shows that the contribution to the growth from
the lateral flux is proportional to $L$ for short NWs. For long NWs,
the adatoms arriving at the middle part of the NW are incorporated
locally (with the formation of a plateau) and the contribution to
the NW axial growth is proportional to $\lambda_{NW}$.

Using Eq.~\ref{ns1} and \ref{P3}, the NW axial growth rate is:

\begin{equation}\label{P4}
\frac{dL}{dt}=\frac{\alpha_L}{\alpha_s}\frac{2\lambda_{NW}^2}{R_0
\lambda_s}V_s\frac{1+\frac{J_{NW}}{J_s}\left(\frac{\lambda_{NW}}{\lambda_s}\right)^2
\left[\cosh(\frac{L+\tilde{\lambda}_s}{\lambda_{NW}})-1\right]}
{\sinh(\frac{L+\tilde{\lambda}_s}{\lambda_{NW}})}
\end{equation}
\\
With the simplifying assumption that $\tilde{\lambda}_s$ is
constant, Eq.~\ref{P4} can be integrated, giving the nanowire length
$L$ as a function of the thickness $h$ of the re-grown layer:
 \begin{equation}\label{LH}
\frac{h}{R_0}=\frac{J_s}{J_{NW}}\frac{1}{2} \ln
\frac{1+\frac{J_{NW}}{J_s}\left(\frac{\lambda_{NW}}{\lambda_s}\right)^2\left[\cosh(\frac{L+\tilde{\lambda}_s}{\lambda_{NW}})-1\right]}
{1+\frac{J_{NW}}{J_s}\left(\frac{\lambda_{NW}}{\lambda_s}\right)^2\left[\cosh(\frac{\tilde{\lambda}_s}{\lambda_{NW}})-1\right]}
\end{equation}

\begin{figure}[h]
 \centering
 \includegraphics[width=\columnwidth]{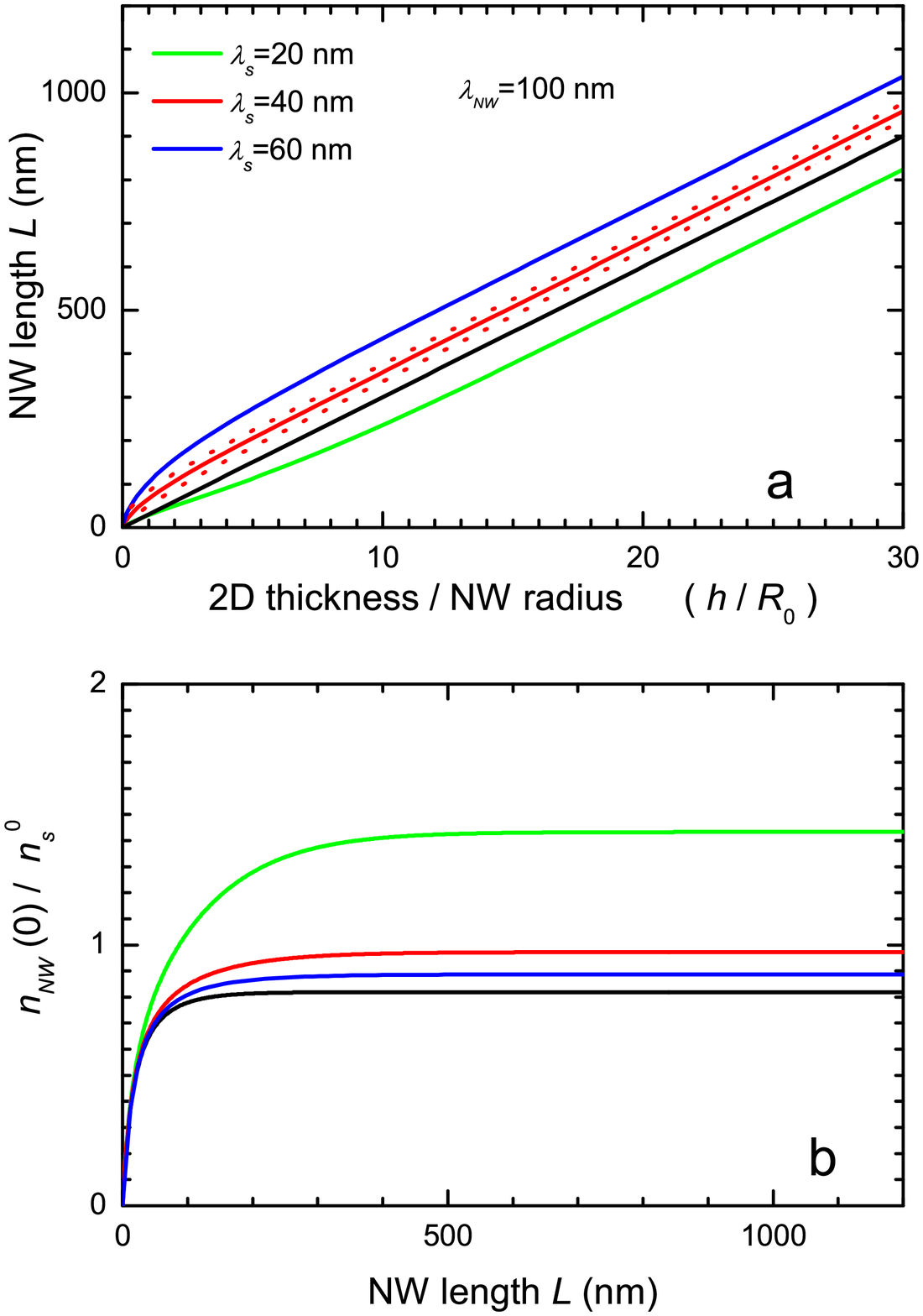}
\caption[]{(a) Nanowire length $L$ as a function of the thickness of
the re-grown layer $h$ divided by the radius of the catalyst
particle $R_0$ (Eq.~\ref{LH}). The continuous curves are plotted
for: $J_{NW}/J_s =0.15$, $\lambda_{NW}= 100 $ nm,
$\tilde{\lambda}_s=20$ nm and three different values of $\lambda_s=
$ as indicated. The dotted red curves have the same parameter except
for $\tilde{\lambda}_s$ changed to 10 nm and 50 nm. Finally, the
black line is the asymptotic line due to lateral flux. (b)
$\omega_R$ as a function of $L$ (Eq.~\ref{P5}). This value is
proportional to the lateral growth rate at the NW base ($z = 0$).
The curves are plotted for: $J_{NW}/J_s =0.15$, $\lambda_{NW}= 200 $
nm, $\tilde{\lambda}_s=20$ nm and three different values of
$\lambda_s= 40, 60, 80$ nm (in red, blue and green). The curve for
the case where there is no contribution from the lateral flux
($J_{NW}^0=0$) is plotted in black.}
  \label{FigA2}
\end{figure}

This dependence of $L$ on $h$ is plotted in Fig.~\ref{FigA2}a for
different contributions from the substrate (different values of
$\lambda_s$).

For the long NWs $L\gg \lambda_{NW}$, the contribution to the NW
axial growth from the adatoms arriving to the substrate and
diffusing up the NW walls up to the catalyst droplets is negligible.
The dominant contribution to the axial growth is that from the
lateral flux, with a constant growth rate $\frac{L}{h}\sim
2\frac{J_{NW} }{J_s} \frac{\lambda_{NW}}{R_0}=2\frac{ \tan \alpha}{
\pi} \frac{\lambda_{NW}}{R_0}$ (dashed black line). A change of the
values of $\tilde{\lambda}_s$ to 10 nm and to 50 nm (extreme values
observed in our samples) makes only minor changes of $L(h)$ (red
dotted lines), which validates the hypothesis of a constant
$\tilde{\lambda}_s$ value.

Hence, the contribution from the substrate influences the axial
growth rate of the NWs only at the early stages of growth. For short
NWs ($L <  \lambda_{NW}$), it determines how fast the NW starts to
grow: the growth speed at $\lambda_s=40$ nm is lower than the final
slope while at $\lambda_s=80$ nm it is larger.

The lateral growth rate at the NW base ($z = 0$), in the case where
$\tilde{\lambda}_s<<\lambda_{NW}$, is given by:
 \begin{eqnarray}\label{P5}
\frac{dR}{dt}&=&\frac{\alpha_{NW}}{\alpha_s}\left(\frac{\lambda_s}{\lambda_{NW}}\right)^2
V_s \: \omega_R(z=0)\\
\text{where}\nonumber\\
 \omega_R(z=0) &=&
\frac{\sinh(\frac{L}{\lambda_{NW}})+\frac{J_{NW}}{J_s}\frac{\lambda_{NW}\tilde{\lambda}_s}{(\lambda_s)^2}\left[\cosh(\frac{L}{\lambda_{NW}})-1\right]}
{\sinh(\frac{L+\tilde{\lambda}_s}{\lambda_{NW}})}\nonumber
\end{eqnarray}

In Fig.~\ref{FigA2}b we plot the value of $\omega_R$, for different
contributions from the substrate (different values of $\lambda_s$).
We observe that for short NWs, the lateral growth is slow and that
the lateral flux does not influence the growth (the four lines are
merged together). For long NWs, the contribution from the lateral
flux remains relatively weak, and the lateral growth rate is
constant with $R/h\sim (\lambda_s / \lambda_{NW})^2$, as a result of
$n_{NW}(z=0)\sim n_s^0$. Thus, the lateral growth at the NW base is
a good measure of the NW growth time, which can be interesting in
samples where the NWs start to grow at different times.

Finally, the variation of the NW radius at the base, $R$, with the
NW length can also be determined:

\begin{equation}\label{P6}
\frac{dR}{dL}=\frac{\alpha_{NW}}{2\alpha_L}\frac{R_0}{\lambda_{NW}}
\frac{\sinh(\frac{L}{\lambda_{NW}})+\frac{J_{NW}}{J_s}\frac{\lambda_{NW}\tilde{\lambda}_s}{(\lambda_s)^2}\left[\cosh(\frac{L}{\lambda_{NW}})-1\right]}
{1+\frac{J_{NW}}{J_s}\left(\frac{\lambda_{NW}}{\lambda_s}\right)^2\left[\cosh(\frac{L+\tilde{\lambda}_s}{\lambda_{NW}})-1\right]}
\end{equation}
which can be integrated to obtain $R(L)$. If $\tilde{\lambda}_s \ll
\lambda_{NW}$, a good approximation is to ignore the second term in
the numerator, so that
\begin{eqnarray}\label{RL}
\frac{R-R(0)}{R_0}=\frac{\alpha_{NW}}{\alpha_L}\frac{1}{2}
\frac{J_s}{J_{NW}}\left(\frac{\lambda_s}{\lambda_{NW}}\right)^2\times\nonumber\\
\ln
\frac{1+\frac{J_{NW}}{J_s}\left(\frac{\lambda_{NW}}{\lambda_s}\right)^2\left[\cosh(\frac{L+\tilde{\lambda}_s}{\lambda_{NW}})-1\right]}
{1+\frac{J_{NW}}{J_s}\left(\frac{\lambda_{NW}}{\lambda_s}\right)^2\left[\cosh(\frac{\tilde{\lambda}_s}{\lambda_{NW}})-1\right]}
\end{eqnarray}
Here we made a simplified distinction between the contact radius
$R_0$ during the main part of the growth, and its initial value
$R(0)$ at the beginning of the NW growth.

\begin{acknowledgments}
This work was performed in the joint CNRS-CEA group
"\emph{Nanophysique \& semiconducteurs}", and in the frame of the
French National Agency project "Magwires" (ANR-11-BS10-013). We
thank all the members of the Magwires project for many discussions.
We also acknowledge a grant from the \emph{Laboratoire d'excellence}
LANEF in Grenoble (ANR-10-LABX-51-01).
\end{acknowledgments}

\end{document}